\documentclass[10pt]{iopart}
\expandafter\let\csname equation*\endcsname\relax
\expandafter\let\csname endequation*\endcsname\relax
\usepackage{graphicx}
\usepackage{lipsum}
\usepackage{bm}
\usepackage{tabu}
\usepackage[utf8]{inputenc}
\usepackage[T1]{fontenc}
\usepackage{mathptmx}
\usepackage{color}
\usepackage{xcolor}
\usepackage{float}
\usepackage{graphicx}
\usepackage{caption}
\usepackage{subcaption} 
\usepackage{cite}
\usepackage{iopams}  
\usepackage{amsmath}
\usepackage{tikz}
\usepackage{bbold}

\newcommand{\sgn}[0]{\text{sgn}}

\begin{document}
\title{Tunable cylindrical vector beam generation via low-cost printed binary holograms}

\author{Emilio E. Ramos-Torres$^1$, Beatriz Morales Cruzado $^{2}$, Benjamin Perez-Garcia$^{3}$, Carmelo Rosales-Guzmán $^{4,*}$}

\address{$^1$ Divisi\'on de Ciencias e Ingenier\'ias, Campus Le\'on, de la Universidad de Guanajuato, Loma del Bosque 103, Col. Lomas del Campestre, 37150, Le\'on, M\'exico}

\address{$^2$ SECIHTI - Centro de Investigaciones en \'Optica, A.C., Loma del Bosque 115, Colonia Lomas del Campestre, C.P. 37150 Le\'on, Guanajuato, Mexico}

\address{$^3$ Photonics and Mathematical Optics Group, Tecnologico de Monterrey, Monterrey 64849, Mexico}

\address{$^4$ Centro de Investigaciones en \'Optica, A.C., Loma del Bosque 115, Colonia Lomas del Campestre, C.P. 37150 Le\'on, Guanajuato, Mexico}

\ead{carmelorosalesg@cio.mx}

\begin{abstract}
We report a low-cost method for generating cylindrical vector beams using binary holograms printed on acetate sheets and a modified Michelson interferometer incorporating a cylindrical-lens mode converter. By simply exchanging the hologram the device produces a variety of CVBs with tunable spatial–polarisation nonseparability. The transverse polarisation distributions reconstructed via Stokes polarimetry show spatial–polarisation features consistent with numerical simulations. The degree of nonseparability is further quantified using the vector quality factor (concurrence), demonstrating values in good agreement with theoretical expectations across the generated states. The use of wave-plate retarders enables continuous tuning from scalar to fully vector beams. The simplicity, robustness, and low cost of the proposed system make it an attractive alternative to programmable modulators for compact optical platforms and teaching laboratories.      
\end{abstract}
 
\ioptwocol
\maketitle
\section{Introduction}

Structured light has become a central theme in modern optics, enabling advances in both fundamental science and applied technologies \cite{Roadmap,Shen2022,rosales2024perspective,Forbes2021StructuredLight}. Within this rapidly expanding field, vector beams-optical fields exhibiting nonseparability between spatial and polarisation degrees of freedom-occupy a particularly important role. For example, their close mathematical analogy to quantum entanglement has motivated the development of concepts such as “classical entanglement’’ \cite{Shen2022,forbes2019classically}, while the topologically rich structures encoded in their Stokes fields have inspired the emerging area of optical skyrmions \cite{chen2025Skyrmions,shen2024skyrmions,shen2025topological}. At the same time, vector beams continue to demonstrate strong practical value in applications including optical trapping, high-capacity communications, optical metrology, and sensing \cite{shvedov2014,rosales2024perspective,BergJohansen2015,Milione2015}. Their versatility across both fundamental and applied domains underscores the importance of developing accessible, high-quality methods for their generation.

Creating vector beams typically requires simultaneous control over spatial and polarisation degrees of freedom. This can be achieved through geometric-phase devices-such as q-plates \cite{Marrucci2006} and metasurfaces \cite{Devlin2017arbitrary}-which imprint spin-dependent phase shifts to generate specific vector states. Alternatively, dynamic-phase approaches employ interferometric arrangements or sequential spatial modulation to independently process orthogonal polarisation components \cite{RodriguezFajardo2024,Maurer2007}. In recent years, programmable Spatial Light Modulators (SLMs) \cite{Neil2009,Moreno2012,Mitchell2017,Rong2014,SPIEbook} and Digital Micromirror Devices (DMDs) \cite{Scholes2019,Gong2014,Hu2022,Ren2015,Lerner2012} have become the dominant tools for this purpose, enabling the realisation of arbitrary vector fields through computer-generated holography. Despite their flexibility, these devices remain expensive, fragile, and often unsuited for compact or field-deployable systems. Their cost also makes them impractical for instructional laboratories, where accessible demonstrations of structured-light concepts are increasingly in demand.

These challenges highlight the need for robust, low-cost methods capable of producing high-quality vector beams without the complexity of programmable modulators. Recent progress in refractive structured-light elements and printed holography has demonstrated promising alternatives \cite{Torres-Leal2023,PlannarOptics}. In this context, simple printed holographic elements offer a compelling route toward scalable, deployable, and cost-effective implementations.

In this manuscript, we demonstrate the generation of cylindrical vector beams (CVBs) using low-cost binary holograms printed on acetate sheets and a modified Michelson interferometer. We show that this approach enables the creation of a broad variety of CVBs with tunable vectorness, ranging continuously from scalar to fully vector beams \cite{Ndagano2016}. The transverse polarisation distributions reconstructed via Stokes polarimetry reveal high-quality spatial–polarisation structures that exhibit good agreement with numerical simulations. We further characterise the degree of nonseparability using the vector quality factor (concurrence $C$), a quantum-inspired metric widely used to quantify vector beam purity \cite{Selyem2019,McLaren2015}. It is worth emphasising that while printed holography and refractive structured-light elements have emerged as low-cost alternatives to programmable modulators, existing demonstrations have been largely limited to the generation of static scalar fields. To the best of our knowledge, no previously reported printed-hologram method provides a mechanism to generate the paired $\pm\ell$ modes with controlled relative phase and orthogonal polarisation required for high-purity cylindrical vector beams. Furthermore, current low-cost solutions lack any means of continuously tuning spatial–polarisation nonseparability. These limitations highlight the need for an accessible approach capable of producing tunable CVBs, thereby motivating the method introduced in this work.

Our results demonstrate that printed binary holograms constitute a practical and reliable alternative to programmable spatial modulators for generating vector beams. Their low cost, simplicity, and ease of integration make them well suited for compact photonic systems and for deployment in undergraduate teaching laboratories, where access to advanced structured-light tools is often limited. This work therefore provides a pathway for democratising vector-beam technologies and extending their reach beyond traditional research settings.

\section{Mathematical background}

\subsection{Light beam shaping printed holograms}
We construct complex optical fields from the solutions of the paraxial wave equation.  In cylindrical coordinates, these solutions correspond to the well-known Laguerre-Gaussian (LG) modes, which at the plane $z=0$ can be expressed as
\begin{equation}
\begin{split}
          LG_p^{\ell}(\rho,\varphi)&=B\left(\frac{\sqrt{2}\rho}{w_0}\right)^{|\ell|}\exp{\left(-\frac{\rho^2}{w_0^2}\right)} \\
          &\times L_p^{|\ell|} \left(\frac{2\rho^2}{w_0^2}\right) \exp{(i\ell\varphi)},
\end{split}
\end{equation}
where $(\rho,\varphi)$ are the polar coordinates, $\ell\in\mathbb{Z}$ is the topological charge, $p\in\mathbb{Z}{\ge 0}$ is the radial index, $w_0$ is the beam waist, $L_p^{|\ell|}(\cdot)$ are the associated Laguerre polynomials of order $(|\ell|,p)$, and $B$ is a normalisation constant.  This complex optical field was encoded in a binary amplitude hologram using the transfer function \cite{lee1974,Lee1979, Mirhosseini2013}
\begin{equation}
    T(\rho,\varphi)=\frac{1}{2}+\frac{1}{2}\sgn\{\cos[p(\rho,\varphi)]+\cos[q(\rho,\varphi)]\}, 
    \label{Hologram1}
\end{equation}
where $\sgn\{\cdot\}$ represents the sign function. The terms $p(\rho,\varphi)$ and $q(\rho,\varphi)$ are encoding functions of the amplitude and phase, respectively,  defined as
\begin{equation}
\begin{aligned}
    p(\rho,\varphi) &= \arcsin\left[\frac{|U_{\ell,p}(\rho,\varphi,)|}{\max\{ |U_{\ell,p}(\rho,\varphi,)|\}} \right],\\
    q(\rho,\varphi) &= \phi(\rho,\varphi)+2\pi(\nu\rho+\eta\varphi).
    \label{Hologram2}
\end{aligned}
\end{equation}
Here, $|U_{\ell,p}|$ and $\phi(\rho,\varphi)$ denote the amplitude and phase of the target field, while the term $2\pi(\nu\rho,\eta\varphi)$ adds a linear phase grating that redirects the encoded $LG_p^{\ell}$ field into the first diffraction order. This spatial-frequency offset separates the desired reconstructed mode from the undiffracted zero and higher diffraction-orders generated by the binarisation process, allowing straightforward spatial filtering with an aperture.

This encoding strategy may be understood as generating a virtual interference pattern between the target complex field and a reference plane wave carrying a known grating frequency. The sum of cosine terms in Eq. \ref{Hologram1} produces a fringe pattern whose local duty cycle represents the amplitude of the target $LG_p^{\ell}$ mode, while the spatial position of the fringes map the phase. The subsequent binarisation by the sign function turns this continuous fringe pattern into a square-wave approximation.

Binary encoding introduces well-known limitations: the diffraction efficiency of amplitude-only holograms is intrinsically lower (typically 8-12$\%$ for the first order), and higher diffraction orders carry residual energy. However, for our application -where low cost, robustness, and simplicity are prioritised- these limitations are acceptable. The first diffraction order, isolated with an aperture, provides a clean reconstruction of the target LG field and is sufficient for generating cylindrical vector beams of good quality in the subsequent interferometric stage.

\begin{figure}[tb]
    \centering
    \includegraphics[width=0.98\columnwidth]{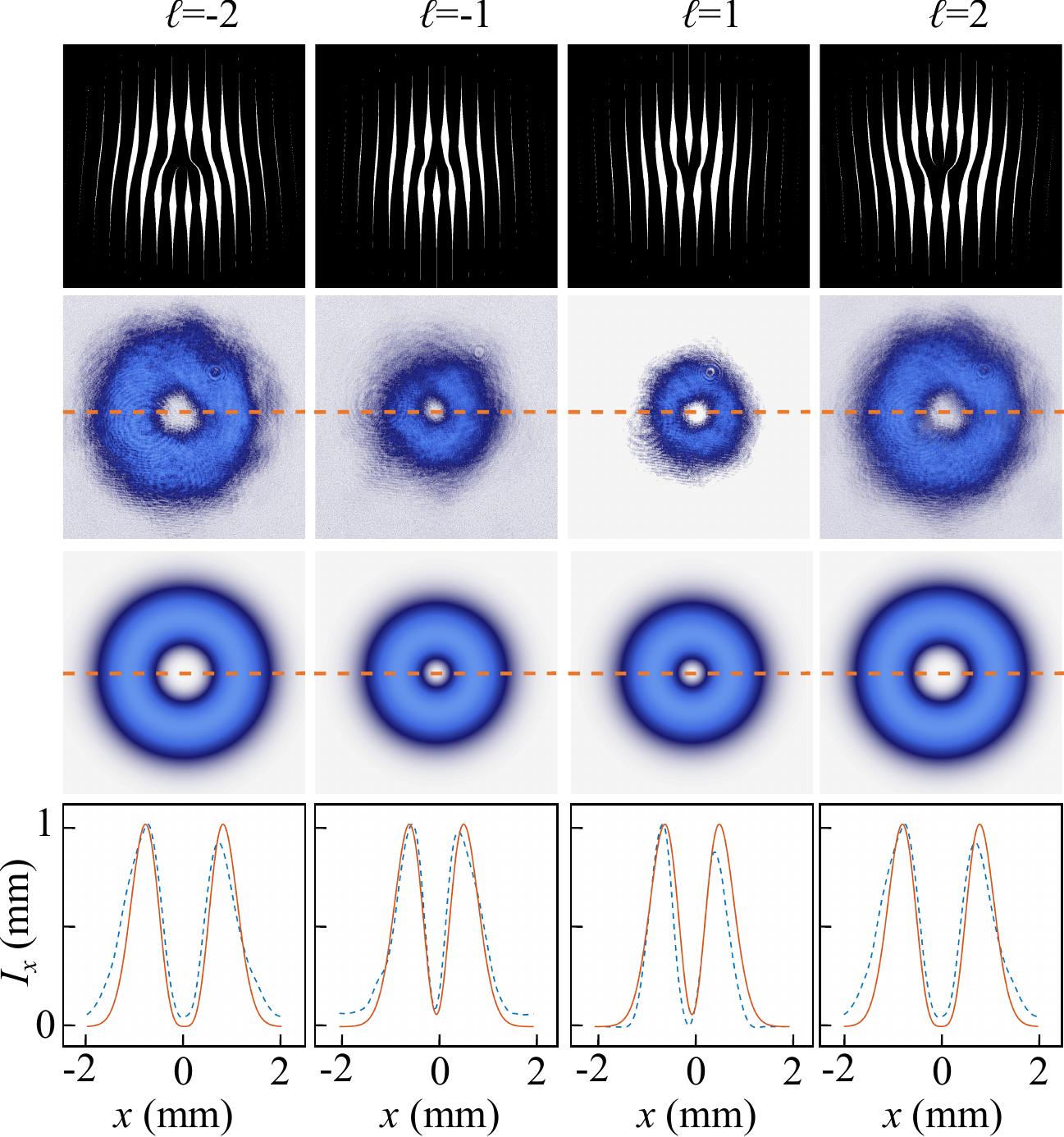}
    \caption{The top row shows the binary amplitude holograms designed for LG beams with topological charges $\ell=-2,-1,+1,+2$. The second and third rows show the experimentally measured intensity distributions of the first diffraction order at $z=0$, together with their corresponding theoretical counterparts. The bottom row shows the horizontal cross-sections (along the dashed line) of the normalised experimental and theoretical intensities, illustrating the good agreement between the beams generated with the printed holograms and the target LG modes.}
    \label{fig:comparisson}
\end{figure}

The binary holograms were fabricated using a high-resolution printer (imagesetter Agfa AVANTRA 44) operating at 2400 Dots Per Inch (DPIs), corresponding to a pixel size of approximately 10 $\mu$m, and printed as positive binary amplitude masks, meaning that regions where $T(\rho,\varphi)=1$ correspond to transparent pixels on the acetate sheet, while regions where $T(\rho,\varphi)=0$ appear opaque. This ensures a direct and physically intuitive mapping between the encoded binary pattern and the transmitted optical field. By way of example, Fig. \ref{fig:comparisson} shows images of the binary holograms for LG beams of topological charges $\ell=\pm 2$ and $\pm 1$ (top row). In a similar way, the intensity profile of the experimentally generated beams compared with their theoretical counterpart are shown in the second and third row, respectively, at the plane $z=0$. The last row shows a one-dimensional cross-section along the horizontal direction, marked with a dashed line, of the normalised intensity. Notice the good agreement of both the experimental and theoretical beams.

\subsection{Cylindrical vector beams}

As it is well known, CVBs can be described mathematically as a nonseparable superposition of the spatial and polarisation degrees of freedom, which for the particular case of LG modes can be expressed as \cite{Galvez2012}
\begin{equation}
\mathbf{U}(\mathbf{r}) = \cos\theta\,  LG_p^{\ell}(\mathbf{r}) \mathbf{\hat{e}_R} + \sin\theta\,  LG_p^{-\ell}(\mathbf{r}) e^{i\delta}\mathbf{\hat{e}_L},
\label{LGVB}
\end{equation}
where the unitary vectors $\mathbf{\hat{e}_R}$ and $\mathbf{\hat{e}_L}$ represent the right and left-handed circular polarisation. The parameter $\theta \in [0, \pi/2]$ is a weighting coefficient that allows to tune the purity or "vectorness" of the generated mode. The parameter $\delta \in [0,2\pi]$ is an intermodal phase that allows a transition between different polarisation states distributions.  

\section{Generation and characterisation of cylindrical vector modes}
\subsection{Experimental setup}
\begin{figure}[tb]
    \centering
    \includegraphics[width=1.0\columnwidth]{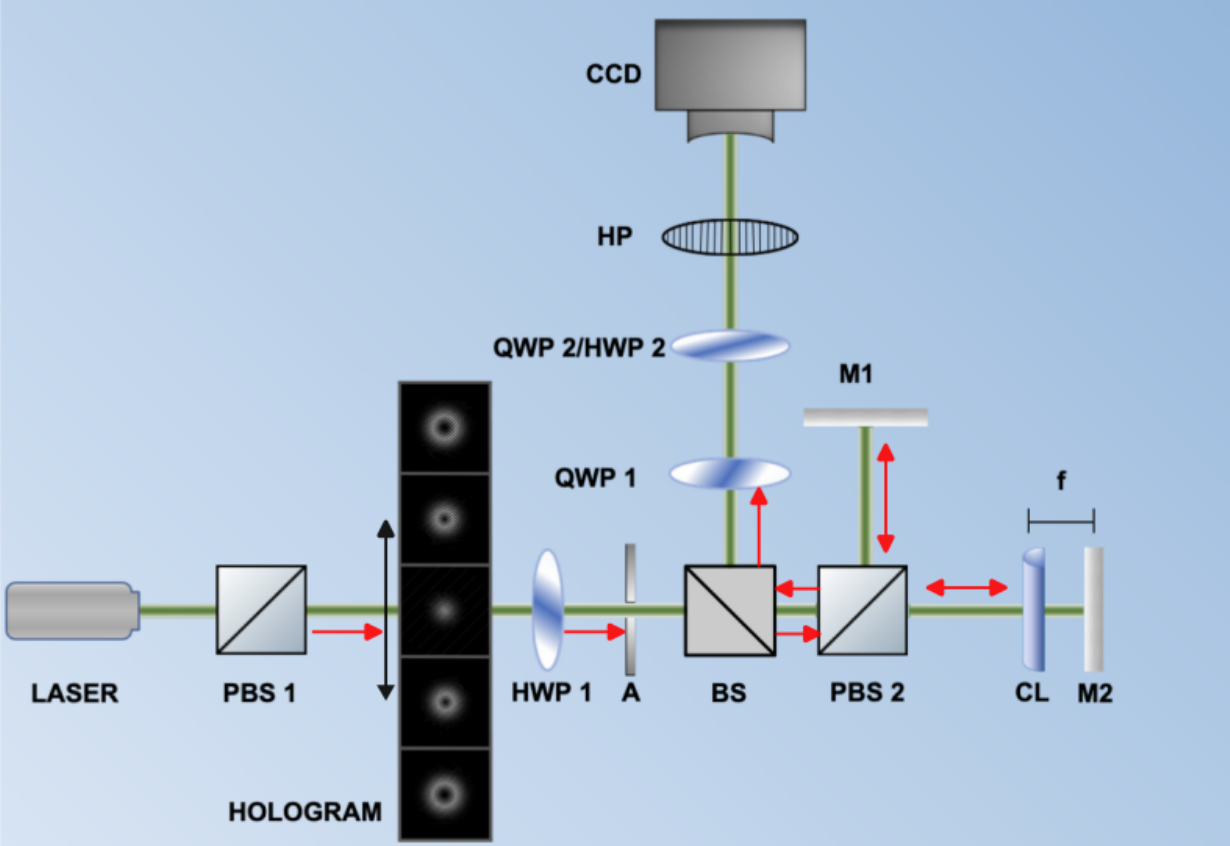}
    \caption{Experimental setup for generating and characterising LG vector beams from a low-cost printed hologram. PBS: polarising beam splitter, BS: beam splitter, HWP: half-wave plate, QWP: quarter-wave plate, CL: cylindrical lens, M: Mirror, HP: Horizontal Polariser, CCD: Charge-coupled device camera.}
    \label{fig:experimental setup}
\end{figure}

The experimental setup implemented to generate LG vector beams relies on a modified Michelson interferometer, as illustrated in figure~\ref{fig:experimental setup}. A collimated laser diode (CPS532, Thorlabs) operating at $\lambda = 532~\mathrm{nm}$ is first sent through a polarising beam splitter (PBS 1) to filter the intrinsic elliptical polarisation of the diode into a well defined horizontal polarisation state. The beam is then passed through a low-cost binary hologram printed on a rectangular acetate sheet which contains a series of LG beams with topological charges ranging from $\ell=-5$ to $\ell=+5$. The hologram is mounted on a well-aligned motorised linear translation stage (X-axis), which allows precise lateral displacement of the hologram so that the beam is directed onto the region corresponding to the desired topological charge. A half-wave plate (HWP 1) is subsequently used to set the required polarisation state, while an aperture diaphragm spatially filters the desired diffraction order corresponding to the selected LG mode. The resulting beam is then directed towards a second PBS (PBS 2), where it is separated into its horizontal and vertical polarisation components. The vertically polarised component is reflected towards a mirror (M1), positioned $4~\mathrm{cm}$ from the PBS 2, whereas the horizontally polarised component propagates through a $\pi$-mode converter. This converter is implemented using a single cylindrical lens (CL) with focal length $f=30~\mathrm{mm}$, placed at a distance $f$ from a reflecting mirror (M2) which is also positioned $4~\mathrm{cm}$ away from the PBS 2. After passing through the cylindrical lens, the beam is reflected by the mirror and propagates back through the same lens, forming a double-pass configuration. This arrangement is equivalent to a system of two identical cylindrical lenses separated by a distance $2f$, fulfilling the condition required for $\pi$-mode conversion. As a result, the sign of the topological charge of the transmitted beam is inverted, changing from $+\ell$ to $-\ell$ or vice versa.  The reflected beams with opposite topological charges and orthogonal linear polarisations are then recombined at the PBS 2 to generate a vectorial beam, which exits through the same input port. A beam splitter (BS) is then used to isolate the generated vector beam which is then transformed from the linear polarisation basis to the circular with a quarter-wave plate (QWP 1). We note that the performance of the $\pi$-mode converter is moderately sensitive to alignment. In our implementation, the most critical parameter is the angular orientation of the cylindrical lens: deviations larger than $0.5\deg$ from the optimal alignment noticeably reduce the overlap between the input and converted modes, leading to incomplete inversion of the topological charge. A secondary sensitivity arises from the longitudinal position of mirror M2, which must remain within approximately $\pm$0.5 mm of the focal plane to preserve the symmetry of the double-pass geometry. Small lateral displacements of the lens or mirror mainly introduce slight distortions in the ring intensity but do not prevent mode inversion. These tolerances provide practical guidance for replicating the setup and ensure reliable mode conversion once the system is initially aligned. The reconstruction of the transverse polarisation distribution was performed through Stokes polarimetry. For this, a series of polarisation projections were implemented using a half-wave plate (HWP 2), a quarter-wave plate  (QWP 2) and a horizontal polariser (HP). The resulting intensity distributions were recorded using a charge-coupled device (CCD) camera (UI-3240CP-C-HQTL, Thorlabs).

It is worth mentioning that the total optical throughput of the system, from the input of the printed hologram to the output of the interferometric stage lies in the range of 1-3\%, which depends on several factors, such as the selected topological charge, the spatial filtering of the first diffraction order, and the cumulative losses introduced by the PBSs and the additional BS used to extract the vector beam from the setup. To achieve higher efficiencies some strategies can be implemented, such as inserting additional QWP inside the Michelson interferometer.

\subsection{Polarisation reconstruction of CVB}

\begin{figure}[tb]
   \centering
    \includegraphics[width=0.49\textwidth]{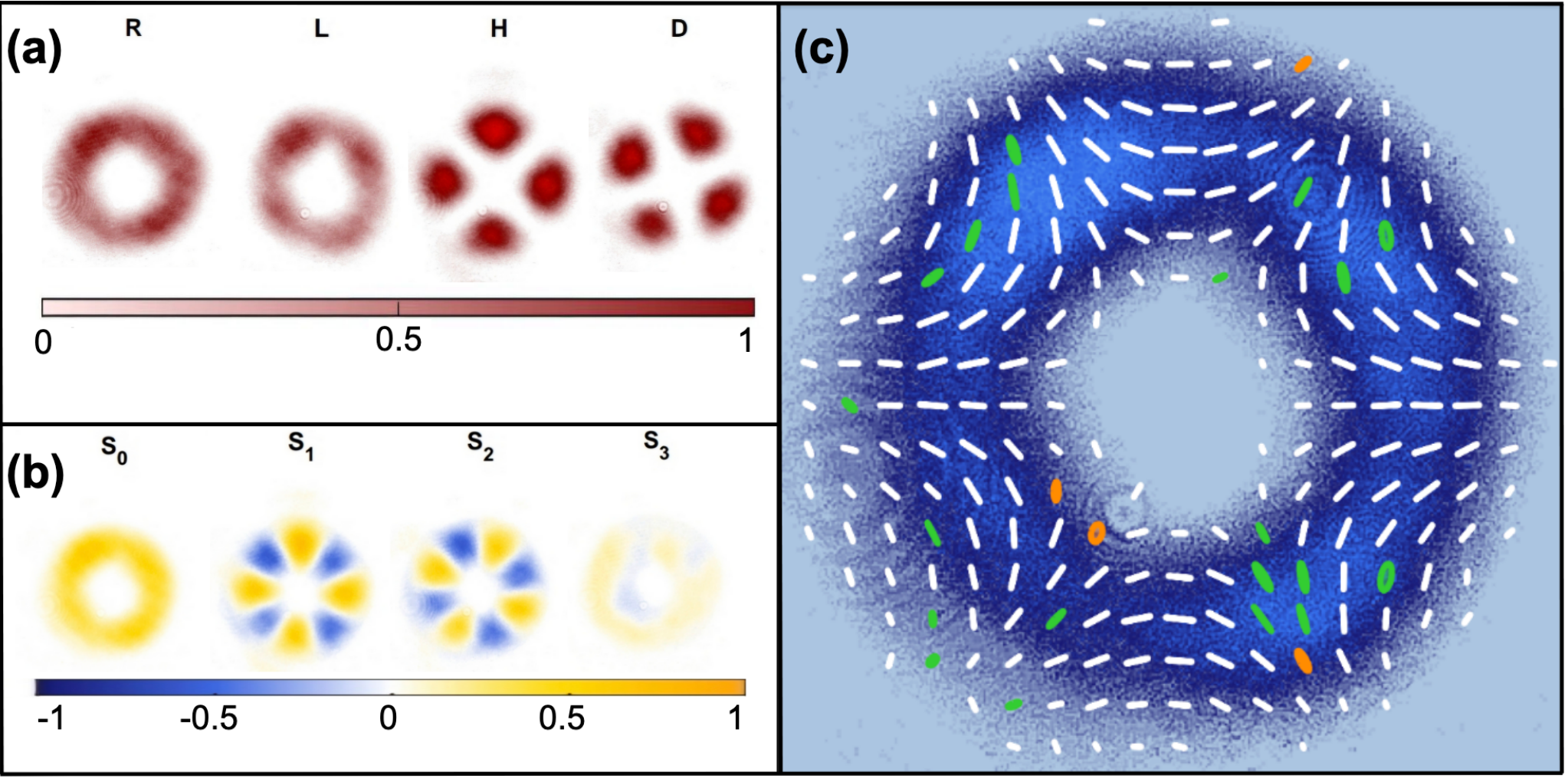}
    \caption{(a) Example of the measured intensity distributions required to compute the Stokes parameters shown in (b) for the CVB given by the parameters $\ell = 2$, $\theta=\pi/4$ and $\delta=0$. The Stokes parameter were normalised to $ S_0$}. (c) Intensity and reconstructed transverse polarisation distribution of the CVB.
\label{fig:stokes-Intensities}  
\end{figure}

As mentioned above, the transverse polarisation distribution of the generated beams was reconstructed using Stokes polarimetry through a series of four intensity measurements\cite{Goldstein2011}. The relation between the required intensities and the Stokes parameters are given by:
\begin{equation}
\begin{aligned}
S_0 &= I_R + I_L, \\
S_1 &= 2I_H - S_0, \\
S_2 &= 2I_D - S_0, \\
S_3 &= I_R - I_L,
\end{aligned}
\label{eq:stokes}
\end{equation}
where $I_H$, $I_D$, $I_R$ and $I_L$ correspond to the measured intensities associated with horizontal, diagonal, right- and left-handed circular polarisation projections, respectively. In the experiment, these polarisation projections are obtained using a combination of phase retarders and a linear polariser. Specifically, the HWP in combination with the HP is employed to measure the horizontal and diagonal intensity components ($I_H$ and $I_D$), while the QWP in combination with the HP is used to obtain the right- and left-handed circular polarisation intensity component ($I_R$ and $I_L$). The resulting intensity distributions were recorded using a CCD camera, and the acquired images are then used to compute the spatially resolved Stokes parameters via equation~\eqref{eq:stokes}. Once the Stokes parameters are determined, the full transverse polarisation distribution of the vector beam can be reconstructed. By way of example Fig.~\ref{fig:stokes-Intensities} (a) shows experimental images of the intensities obtained for a CVB of parameters $\ell=2$, $\theta=\pi/4$ and $\delta=0$. Similarly, Fig.~\ref{fig:stokes-Intensities} (b) shows images of the Stokes parameters reconstructed from these intensities. Finally, Fig.~\ref{fig:stokes-Intensities} (c) shows the intensity profile of the CVB, overlapped with its transverse polarisation distribution, reconstructed from the Stokes parameters \cite{Goldstein2011}.

\subsection{Experimental generation of CVB}

\begin{figure}[tb]
   \centering
    \includegraphics[width=0.49\textwidth]{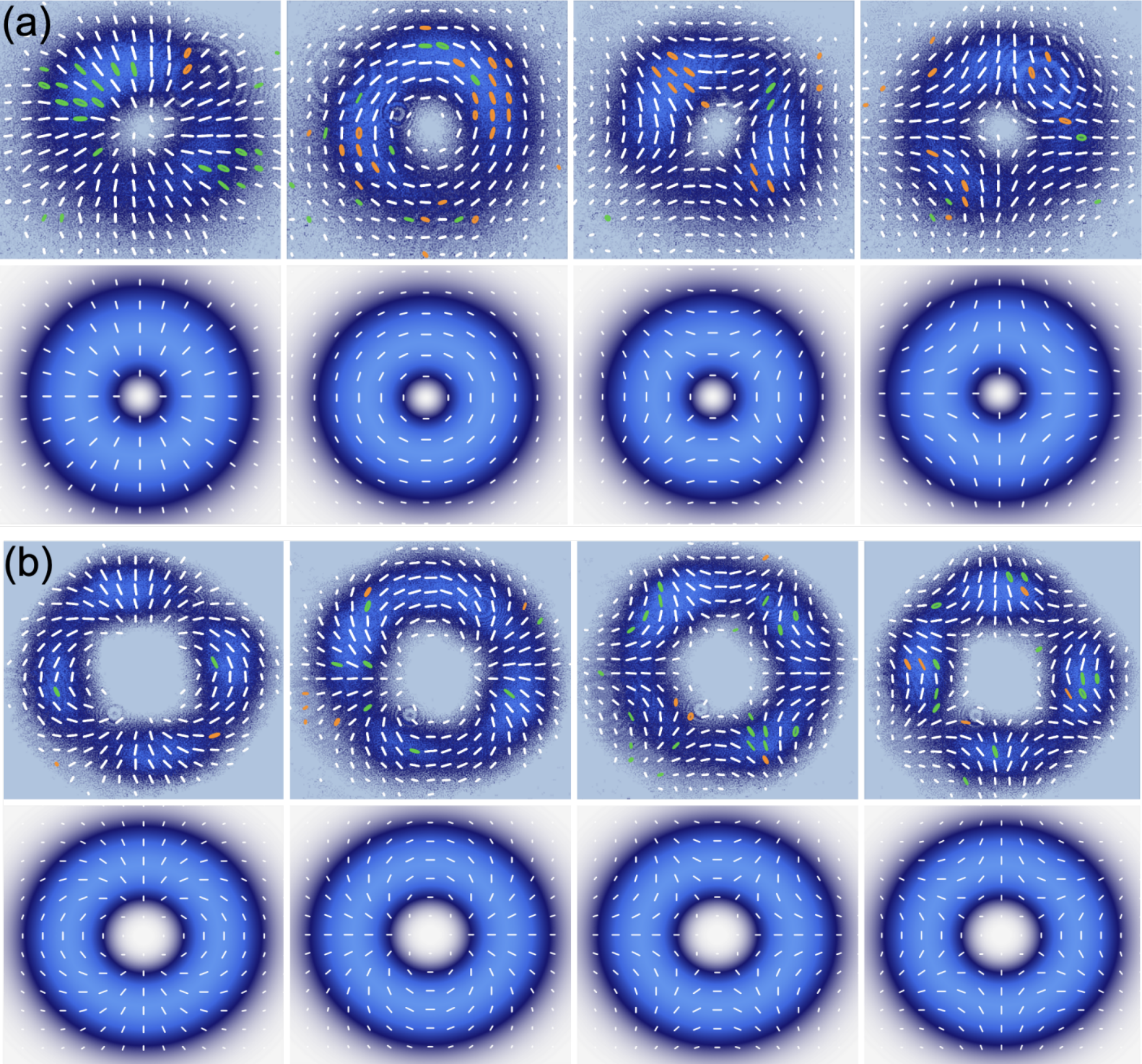}
    \caption{Experimentally reconstructed transverse polarisation distributions (top row) and corresponding numerical simulations (bottom row) for a Laguerre--Gaussian vector beam with topological charge (a)  $\ell = 1$ and (b) $\ell = 2$. All experimental and simulated intensity distributions are normalised to their individual maxima.}
\label{fig:reconstrucciones}
\end{figure}

Representative examples of the generated vector beams, compared to numerical simulations are presented in this section. The numerical simulations were performed using the Kirchhoff's diffraction integral \cite{Goodman1996}. As a first example, Fig.~\ref{fig:reconstrucciones} (a) shows a series of CVB associated to a topological charge $\ell = 1$, experiment on the top row and numerical simulations on the bottom one. More specifically, these modes, from left to right, correspond to, $\{LG_0^{1} \mathbf{\hat{e}_R} + LG_0^{-1} \mathbf{\hat{e}_L}\}$, $\{LG_0^{1} \mathbf{\hat{e}_R} - LG_0^{-1} \mathbf{\hat{e}_L}\}$, $\{LG_0^{-1} \mathbf{\hat{e}_R} + LG_0^{1} \mathbf{\hat{e}_L}\}$ and $\{LG_0^{-1} \mathbf{\hat{e}_R} - LG_0^{1} \mathbf{\hat{e}_L}\}$. Notice that discrepancies in the transverse intensity distributions between the experiment and numerical simulations are observed, which might come from experimental misalignment of the interferometer. Nevertheless, the reconstructed polarisation at each point across the beam, together with the obtained concurrence values, indicates that the experimental reconstruction is in good agreement with the corresponding  simulations.

Figure~\ref{fig:reconstrucciones} (b) shows additional examples of CVB associated with the topological charge $\ell = 2$. More specifically, the corresponding vector beams, from left to right are $\{LG_0^{2} \mathbf{\hat{e}_R} + LG_0^{-2} \mathbf{\hat{e}_L}\}$, $\{LG_0^{2} \mathbf{\hat{e}_R} - LG_0^{-2} \mathbf{\hat{e}_L}\}$, $\{LG_0^{-2} \mathbf{\hat{e}_R} + LG_0^{2} \mathbf{\hat{e}_L}\}$ and $\{LG_0^{-2} \mathbf{\hat{e}_R} - LG_0^{2} \mathbf{\hat{e}_L}\}$. In this case, the measured intensity distributions display a more pronounced circular symmetry compared to the previous example, and the reconstructed polarisation patterns are also consistent with the simulated predictions.

In both cases, the reconstructed beams do not exhibit perfectly uniform intensity around the ring. This behaviour can be attributed to experimental uncertainties during the polarisation measurements, as the experimental setup was highly sensitive to environmental perturbations such as acoustic noise and mechanical vibrations. In particular, the process of switching between QWP2 and HWP2 to obtain the required polarisation projections introduced additional disturbances, which affected the measured intensities to varying degrees. A more detailed inspection indicates that the dominant source of discrepancy arises from small mismatches in the optical path lengths of the two arms of the modified Michelson interferometer. Even sub-wavelength phase deviations accumulate into azimuthal nonuniformities when the $\pm\ell$ components are recombined. A secondary contribution comes from angular misalignment in the cylindrical-lens $\pi$-mode converter, where slight deviations from the ideal double-pass geometry alter the modal overlap between the converted and unconverted beams. We also note that a small tilt at PBS2 during recombination can introduce a residual spatial walk-off between the horizontal and vertical components, reducing the circular symmetry of the measured intensity patterns. Future refinements could include implementing shearing-interferometry tests to verify wavefront parity between the two arms, active optimisation of mode matching at the $\pi$-converter, and improved mechanical isolation to minimise environmental perturbations during Stokes projections. These procedures would help further reduce the observed discrepancies and improve reproducibility of the reconstructed vector fields.

\subsection{Tuning the vectorness of CVB}

As stated earlier, the parameter $\theta \in [0, \pi/2]$ allows to tune the coupling strength between the spatial and polarisation degrees of freedom, also known as "vectorness", of the generated vector modes in a continuous way. More precisely, for $\theta=0$ and $\theta=\pi/2$ we obtain scalar beams, whereas for $\theta=\pi/4$, we obtain vector beams. In our experiment, this parameter can be easily changed by rotating HWP 1 in the interval [0, $\pi/2$].

Importantly, while the transverse polarisation structure of a beam provides a qualitative indication of whether a given mode exhibits scalar or vectorial behaviour, it does not provide a measure of its vectorness. Researchers have adapted the concurrence $C$ from quantum mechanics to the field of vector beams, where it has been termed Vector Quality Factor (VQF), to quantify the vectorness \cite{McLaren2015, Ndagano2016}. The VQF takes values in the interval $[0,1]$, where $\mathrm{VQF} = 0$ corresponds to a purely scalar mode with uniform polarisation across the transverse plane, while $\mathrm{VQF} = 1$ indicates a fully vectorial mode with maximal non-separability between spatial and polarisation components. For CVBs, the VQF is related to the parameter $\theta$ as \cite{Ndagano2016}
\begin{equation}
\mathrm{VQF}(\theta) = |\sin(2\theta)|,
\end{equation}
where $\theta$ is the angle of the HWP with respect to the horizontal.

Experimentally, the VQF can be calculated directly from the Stokes parameters according to \cite{Selyem2019}
\begin{equation}
\mathrm{VQF} = \sqrt{1 - \left(\frac{\mathbb{S}_1}{\mathbb{S}_0}\right)^2 - \left(\frac{\mathbb{S}_2}{\mathbb{S}_0}\right)^2 - \left(\frac{\mathbb{S}_3}{\mathbb{S}_0}\right)^2},
\label{eq:concurrence}
\end{equation}
where $\mathbb{S}_0$, $\mathbb{S}_1$, $\mathbb{S}_2$, and $\mathbb{S}_3$ represent the Stokes parameters integrated over the full transverse profile of the beam, i.e, $\mathbb{S}_i=\iint_{-\infty}^\infty S_{i} dA$, for $i=0,1,2,3$.

\begin{figure}[ht!]
   \centering
    \includegraphics[width=0.49\textwidth]{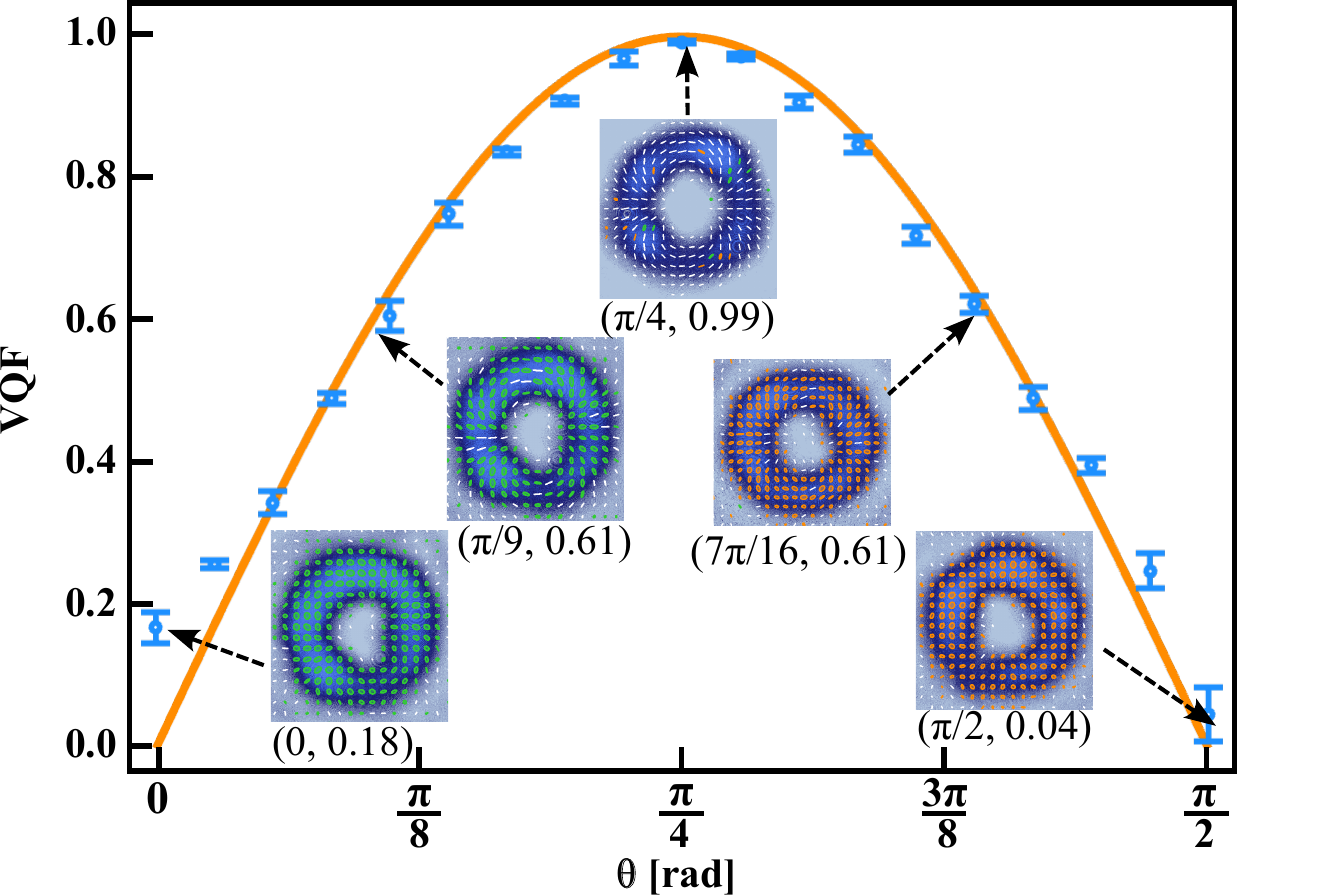}
    \caption{(a) The solid line corresponds to the theoretical values of VQF as a function of $\theta$, whereas blue markers correspond to the mean of four repeated measurements for each setting of the half-wave plate HWP 1. The error bars represent the standard deviation across these measurements. The horizontal axis is explicitly labelled in terms of the parameter $\theta$, which directly corresponds to the rotation angle of HWP 1 in the experimental setup. The insets show the reconstructed polarisation distributions for selected points along the VQF curve. All polarisation maps shown in the inset are normalised to their maximum intensity.}
  \label{VQF}
\end{figure}

We corroborated that our low-cost system can generate CVBs with tuneable vectorness and measured the VQF of the generated beams.  These results are summarised in Fig.~\ref{VQF}, which shows a plot of the VQF as a function of the parameter $\theta$. As mentioned earlier, in the experiment, the vectorness was tuned by rotating HWP 1 from 0 to $\pi/2$, prior to a previous calibration of the same. In total, 19 polarisation settings were measured, with a $5^\circ$ increase between successive points. For each polarisation setting, the measurement was repeated four times in order to estimate the associated experimental uncertainty. The resulting experimental data points are found to be in close agreement with the corresponding theoretical VQF curve. For the sake of clarity, the insets in Fig. \ref{VQF} shows five examples of the transverse polarisation distribution overlapped with the transverse intensity profile for the specific cases $\theta=0$, $\theta=\pi/9$, $\theta=\pi/4$ $\theta=7\pi/16$ and $\theta=\pi/2$, with corresponding VQF values $0.18, 0.61, 0.99, 0.61$ and 0.04, as explicitly labelled in bottom of each inset.  

It is worth mentioning that although the experimental setup was sensitive to environmental perturbations, we obtained a quantitative estimate of the resulting variability through repeated Stokes-parameter measurements. For each value of the parameter $\theta$, the full set of polarisation projections was acquired four times, and the corresponding concurrence values were computed independently. The error bars shown in Fig. \ref{VQF} represent the standard deviation across these four measurements, providing a direct indication of the uncertainty introduced by mechanical vibrations and residual misalignment. While a full root-mean-square analysis of the spatially resolved Stokes parameters is beyond the scope of this work, the relatively small spread of the measured VQF values suggests that the impact of these perturbations on the reconstructed polarisation patterns is limited and does not alter the qualitative agreement with the theoretical predictions.

\section{Discussion}
\label{Discussion}

We have demonstrated that cylindrical vector beams can be generated without relying on costly programmable devices such as SLMs or DMDs, providing a practical route toward democratising structured-light technologies in scenarios where dynamic reconfiguration is not required. Using low-cost printed binary holograms, we generated a variety of cylindrical vector modes, reconstructed their transverse polarisation distributions, and quantified their degree of nonseparability through the VQF. The reconstructed polarisation patterns exhibit spatial-polarisation structures in good qualitative agreement with numerical simulations, while the measured VQF values closely follow the expected theoretical behaviour. Furthermore, the proposed system enables continuous tuning of vectorness, an increasingly relevant capability in emerging applications such as optical communications \cite{Singh2023}.

The simplicity, robustness, and low cost of the setup make it particularly attractive for undergraduate teaching laboratories, where it enables hands-on exploration of advanced structured-light concepts without requiring expensive optical instrumentation. Beyond educational settings, the approach also offers potential for compact and scalable photonic platforms requiring stable vector beams. Although the method is inherently less flexible than programmable modulators—since changing the generated mode requires replacing the printed hologram—it offers several practical advantages, including low cost, immunity to electronic noise, mechanical robustness, and high optical damage tolerance.

Importantly, unlike previous printed-holography and refractive structured-light approaches that have primarily focused on static scalar beam generation, our method enables the generation of tunable cylindrical vector beams through the integration of printed binary holograms with a $\pi$-mode converter in a modified Michelson interferometer. To the best of our knowledge, this constitutes the first low-cost implementation capable of generating paired $\pm \ell$ modes with controlled relative phase and orthogonal polarisation, enabling the synthesis of high-purity CVBs with continuously tunable vectorness through simple wave-plate control.

Although all demonstrations presented here employ modes with $p=0$, the method is readily extendable to higher-order radial modes. The hologram-encoding procedure described in Eqs.~\ref{Hologram1}-\ref{Hologram2} supports arbitrary $(\ell,p)$ combinations, while the $\pi$-mode converter acts exclusively on the azimuthal degree of freedom, preserving the radial structure of the beam. Consequently, CVBs with $p>0$ can be generated without modification of the experimental setup by simply replacing the printed hologram.

Finally, the tunable vector fields produced by this platform may also provide a convenient framework for investigating the behaviour of polarisation singularities under reflection. Recent studies have shown that C-points undergo characteristic topological transformations upon reflection \cite{zhen2025cpoint}. Although such investigations lie beyond the scope of the present work, the proposed platform offers a straightforward and accessible route for exploring these reflection-induced phenomena in future studies.

\section*{Funding}
C.R.G. acknowledges financial support from SECIHTI, México, through the project CBF-2025-I-1804.

\section*{References}
\bibliographystyle{iopart-num}
\bibliography{References}
\end{document}